\documentclass[9.5pt]{article}
\usepackage[preprint]{spconf,amsmath,graphicx}

\usepackage{color}
\usepackage{booktabs}
\usepackage{hyperref}
\hypersetup{
    colorlinks=true,
    linkcolor=blue,
    filecolor=magenta,      
    urlcolor=cyan,
}
 
\newcommand{\parjump}{\vspace{+0.5em}} %{\vspace{+0.5em}}

\title{Libri-Adapt: A new Speech Dataset for Unsupervised Domain Adaptation}

\name{Akhil Mathur$^{\dagger\star}$ \qquad Fahim Kawsar$^{\star}$ \qquad Nadia Berthouze$^{\dagger}$ \qquad Nicholas D. Lane$^{\ddagger}$}
			\address{$^{\dagger}$ University College London, United Kingdom \\
			    $^{\star}$Nokia Bell Labs Cambridge, United Kingdom\\
    		    $^{\ddagger}$University of Oxford, United Kingdom
			    }

\begin{document}
\ninept
\maketitle

% \copyrightnotice{Copyright 2020 IEEE. Published in the IEEE 2020 International Conference on Acoustics, Speech, and Signal Processing (ICASSP 2020), scheduled for 4-9 May, 2020, in Barcelona, Spain. Personal use of this material is permitted. However, permission to reprint/republish this material for advertising or promotional purposes or for creating new collective works for resale or redistribution to servers or lists, or to reuse any copyrighted component of this work in other works, must be obtained from the IEEE. Contact: Manager, Copyrights and Permissions / IEEE Service Center / 445 Hoes Lane / P.O. Box 1331 / Piscataway, NJ 08855-1331, USA. Telephone: + Intl. 908-562-3966.}

\begin{abstract}
% This paper introduces a new dataset, Libri-Adapt, to support unsupervised domain adaptation research on speech recognition models. Built on top of the LibriSpeech corpus, Libri-Adapt contains 7200 hours of English speech divided across 72 domains. Libri-Adapt is primarily targeted for developing ASR models for wearable and embedded devices and is recorded on six different embedded-scale microphones. The domains contained in the dataset are representative of the challenging practical scenarios encountered by ASR models such as microphone variability, speaker accent variability and acoustic environment variability. We conclude by providing a range of baseline results quantifying the impact of various domain shifts on the performance of ASR models.

This paper introduces a new dataset, Libri-Adapt, to support unsupervised domain adaptation research on speech recognition models. Built on top of the LibriSpeech corpus, Libri-Adapt contains 7200 hours of English speech recorded on mobile and embedded-scale microphones, and spans 72 different domains that are representative of the challenging practical scenarios encountered by ASR models. More specifically, Libri-Adapt facilitates the study of domain shifts in ASR models caused by a) different acoustic environments, b) variations in speaker accents, c) previously unexplored factors such as heterogeneity in the hardware and platform software of the microphones, and d) a combination of the aforementioned three shifts. We also provide a number of baseline results quantifying the impact of these domain shifts on the Mozilla DeepSpeech2 ASR model.

\end{abstract}
%
% \vspace{-0.2cm}
\section{Introduction}
% \vspace{-0.2cm}
\label{sec:intro}
Recent advancements in deep neural networks have led to significant improvements in the performance of automatic speech recognition (ASR). However, it is also known that minor deviations between the training and test domains (known as \emph{domain shift}) can significantly degrade the generalizability of deep learning models~\cite{tzeng2017adversarial}. In the case of speech recognition, such perturbations could be introduced in the data by several real-world factors including ambient noise~\cite{barker2015third}, variations in acoustic environments~\cite{vincent2017analysis, barker2018fifth}, accents of speakers~\cite{iskra2002speecon}, or even the microphones used to record the speech~\cite{mathur2019mic2mic, hasca2018} -- and they represent a major challenge to the generalizability of ASR models in unconstrained settings. Clearly, the collection of labeled training data across all these heterogeneous scenarios is prohibitively expensive, therefore \emph{unsupervised domain adaptation} has emerged as a promising technique towards adapting deep neural networks to different but related domains, even in the absence of labeled data from the target domain~\cite{daume2006domain}. 

There are plethora of datasets to study and develop unsupervised domain adaptation techniques for visual recognition (e.g., ~\cite{hoffman2018cycada, tzeng2017adversarial, flexadapt}), however in speech recognition research, there is still a lack of large-scale realistic datasets which contains examples of various forms of \emph{domain shifts} that are expected in real-world scenarios. In this work, we present a new dataset named \emph{Libri-Adapt}, primarily developed to study unsupervised domain adaptation, which contains domain shifts caused by a) variability in recording microphones, b) variability in speaker accents and c) variability in acoustic environments. Libri-Adapt is mainly recorded on consumer-grade, embedded-scale microphones (please refer Table~\ref{tab:mics}) that are becoming popular due to their easy-integration capabilities with microcomputers such as the Raspberry Pi to enable developers to create speech applications for wearable and embedded devices. 

Libri-Adapt is built on top of the Librispeech-clean-100 dataset~\cite{panayotov2015librispeech} which contains 100 hours of US English speech from users reading public-domain audio books. In Libri-Adapt, we record the Librispeech-clean-100 training corpus on six different microphones (five are embedded-scale or smartphone microphones and one is a Shure microphone to establish a higher-quality baseline), in three different accents (US English, British English, and Indian English) and under four synthesized background noise conditions (Clean, Rain, Wind, Laughter). As such, Libri-Adapt has a total of 7,200 hours (6 microphones x 3 accents x 4 environments x 100 hours) of data for training ASR models and can be used to study domain adaptation for potentially 72 different domains (6 microphones x 3 accents x 4 environments). For each domain, we also collect a 5.4 hour held-out test set derived from Librispeech test-clean corpus~\cite{panayotov2015librispeech}. 

In \S\ref{sec:setup}, we describe the methodology used for collecting the dataset followed by an overview of how the dataset is structured. In \S\ref{sec:uda}, we demonstrate the potential of doing unsupervised domain adaptation research on this dataset by training a Mozilla DeepSpeech2~\cite{deepspeech} ASR model on each domain in the dataset and testing it on another domains. Our results show that even state-of-the-art ASR models such as DeepSpeech2 suffer accuracy degradation due to domain shifts -- e.g., the word error rate of a model trained on the Shure microphone increases from 10.19\% (when it is tested on the same microphone) to 24.95\% when it is tested on a Matrix Voice microphone. 

In summary, Libri-Adapt is the largest public dataset targeted for consumer-grade embedded-scale microphones to facilitate the study and development of robust ASR models using domain adaptation techniques. The dataset is released publicly under the CC BY 4.0 license\footnote{https://creativecommons.org/licenses/by/4.0/} and can be accessed from \url{https://bit.ly/2UK1McG}\footnote{Updated baseline results with a newer release of DeepSpeech2 model (0.8.2) are also available at this URL.}.

% \vspace{-0.2cm}
\section{Related Work}
% \vspace{-0.2cm}
\label{sec:rw}
Our work follows a rich tradition of making speech data public to support research on speech and audio recognition. For example, Gemmeke et al.~\cite{gemmeke2017audio} released AudioSet at ICASSP 2017 for the task of audio event recognition. Similarly, a number of speech and keyword datasets have been released, including LibriSpeech~\cite{panayotov2015librispeech}, Common Voices~\cite{commonvoice} and Speech Commands~\cite{warden2018speech}. Researchers have also developed simulated datasets focused on the major challenges in ASR, e.g., handling reverberations~\cite{kinoshita2013reverb, ravanelli2016realistic}, noisy speech~\cite{barker2015third}, speech separation and distant ASR~\cite{ravanelli2016realistic, barker2015third}. A prominent example has been the CHiME challenge~\cite{barker2015third, vincent2017analysis, barker2018fifth} which has released datasets containing real and simulated multi-channel speech data in noisy real-world conditions as well as in multi-speaker scenarios. For the purpose of studying accent variability, prior works have used the Speecon~\cite{iskra2002speecon} family of speech corpora recorded in multiple languages.   

To the best of our knowledge, no prior work has developed a large-scale speech corpus for embedded-scale microphones (refer Table ~\ref{tab:mics} for details) that are gaining popularity nowadays. Further, recent research~\cite{mathur2019mic2mic} has shown that embedded microphones exhibit significant differences in their outputs owing to the variations in their hardware and on-board signal processing pipelines. As such, the primary goal of this paper is to collect speech datasets -- from embedded-scale microphones -- that are large enough to support training of deep ASR models. The variability across microphones can be interpreted as a form of domain shift, thus allowing for development and evaluation of unsupervised domain adaptation algorithms on this dataset. In addition, we are also interested in understanding how microphone variations interact with other challenges for ASR, namely variations in speaker accents and acoustic environments -- as such, we record speech data from 3 speaker accents in 4 acoustic environments on 6 different embedded microphones. 

\parjump
\noindent
\textbf{Unsupervised Domain Adaptation (UDA).} At a high-level, the idea behind UDA is as follows: given a pre-trained deep classifier for a source domain and an unlabeled dataset from a target domain, how can we adapt the weights of the source model such that it shows better performance in the target domain. This class of algorithms has been used recently to adapt ASR models from clean to noisy speech, for speech dereverberation and for speaker gender adaptation~\cite{michelsanti2017conditional, hosseini2018multi, wang2018investigating}. Our work aims to advance this line of research by introducing the large-scale multi-domain Libri-Adapt dataset. 

% \vspace{-0.2cm}
\section{Methodology and Dataset Overview}
% \vspace{-0.2cm}
\label{sec:setup}
In this section, we describe the different domains in the Libri-Adapt dataset and the methodology adopted to create them. 

\parjump
\noindent
\textbf{Microphones.} One of the important goals for Libri-Adapt is to support ASR research on consumer-grade embedded-scale microphones. In the past few years, a number of embedded-scale microphones (both single- and multi-channel) have been released which can be used in conjunction with microcomputers such as Raspberry Pi and cloud-based speech models to develop custom speech analysis systems~\cite{matrixalexa}. However, research~\cite{mathur2019mic2mic} has shown that these microphones could potentially have significant variations in their hardware and software processing pipelines, which in turn could induce domain shift in the speech data obtained from them. Therefore, in order to develop robust ASR systems, we believe it is critical to collect speech datasets from various microphones and evaluate the generalizability of ASR models across them. In this vein, we collect training data (derived from Librispeech-clean-100) from six different microphones as listed in Table~\ref{tab:mics}. Matrix Voice\footnote{https://www.matrix.one/products/voice} and ReSpeaker\footnote{https://respeaker.io/usb\_6+1\_mic\_array/} are circular 7-channel microphone arrays that include on-device acoustic processing algorithms for de-reverberation and delay-and-sum beamforming to combine the outputs of the different channels. PlayStation-Eye is a digital camera used in Sony PlayStation gaming consoles and consists of a 4-channel microphone array used for speech user interaction with the gaming console. Next, we use a Google Nexus 6 as a representative smartphone microphone with which users are likely to interact while using speech applications. Further, we use a single-channel USB microphone as an example of low-end microphones popularly used with Raspberry Pi. Finally, to establish a higher-quality microphone baseline, we employ a Shure MV5 Condenser microphone\footnote{https://www.shure.com/en-GB/products/microphones/mv5/} which can also be used as a plug-and-play microphone with mobile and wearable devices. 

We use a Raspberry Pi as the host device for Matrix, ReSpeaker, USB and PlayStation Eye, while the Shure microphone uses a Macbook Pro as the host device. Each microphone in the dataset can be considered as a unique domain, which allows researchers to perform a variety of domain adaptation experiments. 

% with capabilities voice location tracking, echo cancellation, and background noise suppression. 
% \begin{table}[]
% \centering
% \small
% \begin{tabular}[b]{|c|c|c|p{2cm}|} \hline
% \textbf{Device} & \textbf{Channels} & \textbf{Sampling} & \textbf{Advertised Capabilities} \\ \hline
% Matrix Voice & 2 & \$55 & bla bla \\\hline
% ReSpeaker  & 2 & \$80 & bla bla \\\hline
% USB  & 2 & \$5 & bla bla \\\hline
% \end{tabular}
%     \caption{Caption}
%     \label{tab:mics}
% \end{table}

% \usepackage{booktabs}
\begin{table}[]
\centering
\small
\begin{tabular}{@{}cccp{2.5cm}@{}}
\toprule
\textbf{Device} & \textbf{Channels} & \textbf{\begin{tabular}[c]{@{}l@{}}Maximum \\ Sampling \\ Rate (kHz)\end{tabular}} & \textbf{\begin{tabular}[c]{@{}l@{}}Advertised Signal \\ Processing \\ Capabilities\end{tabular}} \\ \midrule
\begin{tabular}[c]{@{}c@{}}Matrix\\ Voice\end{tabular} & 7 & 8 to 96 & Beamforming, dereverberation, noise cancellation \\ \midrule
Respeaker & 7 & 8 to 48 & Voice activity detection, beamforming, background noise suppression \\ \midrule
\begin{tabular}[c]{@{}c@{}}PlayStation\\ Eye\end{tabular} & 4 & 8 to 48 & Voice location tracking, echo cancellation, background noise suppression.\\ \midrule
USB Mic & 1 & 8 to 48 & - \\ \midrule
Google Nexus 6 & 3 & 8 to 48 & - \\ \midrule
Shure MV5 & 1 & 8 to 48 & Auto gain, Equalization \\ \bottomrule
\end{tabular}
\caption{Technical specification of the microphones used in the Libri-Adapt dataset.}
\label{tab:mics}
\end{table}

\parjump
\noindent
\textbf{Speaker Accents.} Libri-Adapt consists of speech data in three English accents, namely US English, British English and Indian English. To facilitate domain adaptation experiments for countering the domain shift due to variations in speaker accents, we made a design decision to create accented speech that share the same underlying text corpus. In other words, the accented speech subsets have exactly the same sentences being spoken, but in different English accents. For US English, we use the original Librispeech-clean-100 dataset that was recorded with 251 US speakers. To create its variants in British English and Indian English accents, we employed the text-to-speech (TTS) API~\footnote{https://cloud.google.com/text-to-speech/} from Google Cloud Platform which uses a pretrained WaveNet model to generate accented speech in multiple voices. More specifically, we passed the transcripts of the Librispeech-clean-100 corpus to the Google TTS API along with the target accent (British and Indian) and obtained two accented versions of the Librispeech-clean-100 dataset, while ensuring that the underlying spoken sentences remain same in each accent. 

\parjump
\noindent
\textbf{Data Recording Setup.} We adopted the methodology of replay-and-record to collect the Libri-Adapt dataset. Each accented dataset (US English, British English and Indian English) was played on a JBL LSR305 monitor speaker connected to a laptop, and recorded simultaneously on all six microphones in a quiet room. The microphones were kept at a distance of 15cm from the speaker. To ensure that the recordings are time-synchronized across microphones, we implemented a network-based synchronization scheme based on MQTT~\cite{mqtt} messaging protocol. Before a speech file is played, the laptop sends a MQTT message to the recording device hosting each microphone to initiate an audio recording process for the duration of the speech file. Thereafter, the speech file is played on the JBL LSR305 monitor speaker and is recorded simultaneously by all six microphones. After the recording completes, each device sends a confirmation message to the laptop, which proceeds to play the next file in the dataset. 

Indeed, this approach of replay-and-record has its pros and cons. It allows for collecting large-scale datasets (recall that Libri-Adapt has 7200 hours of data) with various domain shifts cheaper than recruiting human subjects -- however on the downside, as the speech is replayed through a speaker, it is impacted by the transfer function of the speaker. To counter this issue, we use a high-quality monitor speaker with a reasonably flat frequency response to replay the audios, which ensures that there is minimal impact on the replay audio quality due to the speaker. More importantly, we argue that the replayed dataset still allows for studying the domain shift problem caused by different microphones because the replay process could be interpreted as a form of covariate shift on the input distribution, that is shared across all microphones. 

\parjump
\noindent
\textbf{Ambient Noise Augmentation.} Finally, we augment the speech datasets collected from all microphones with three types of background noise, namely  `Wind', `Rain' and `Laughter'. To this end, we use the audio samples for `Wind', `Rain' and `Laughter' from the Environmental Sound Classification (ESC-50)~\cite{piczak2015esc} dataset and record them on each of the microphone following our replay-and-record methodology. In total, we record 20 noise samples for each noise type on each microphone. 
Finally, to create the noisy variants of the dataset, each speech file in the original (Clean) dataset is overlaid with a randomly picked noise sample from the same microphone (at SNR of 20dB). As such, we end up with 4 background noise conditions in the dataset: Clean, Rain, Wind, Laughter.  

\parjump
\noindent
\textbf{Dataset Overview.}
Table~\ref{tab:overview} provides an overview of the Libri-Adapt dataset. In total, the dataset has 6 microphone domains, 3 accent domains and 4 background noise domains, resulting in a total of 6 x 3 x 4 = 72 domains. Each domain consists of 28,540 speech files (sampling rate 16KHz) with a duration of 100.6 hours as labeled training data, and 2600 speech files with a duration of 5.4 hours as labeled test data. The dataset facilitates a wide variety of domain adaptation experiments, both supervised and unsupervised. For example, one could train an ASR model with \{\emph{US English} speakers on a \emph{Nexus 6} microphone in \emph{Clean} background noise\} and test its performance on {\{\emph{Indian English} speakers using a \emph{ReSpeaker} microphone with \emph{Rain} background noise\}}.

\begin{table}[]
\centering
\small
\begin{tabular}{@{}llp{4cm}@{}}
\toprule
\textbf{Domain Types} & \textbf{Count} & \textbf{Domain Names} \\ \midrule
Microphones & 6 & Matrix Voice, ReSpeaker, PlayStation Eye, USB, Shure, Nexus 6 \\
Accents & 3 & US English, Indian English, British English \\
Background Noise & 4 & Clean (or None), Rain, Wind, Laughter \\ \bottomrule
\end{tabular}
\caption{Overview of the Libri-Adapt dataset}
\label{tab:overview}
\end{table}

% \vspace{-0.2cm}
\section{Effect of Domain Shift on ASR Models}
% \vspace{-0.2cm}
\label{sec:uda}
In this section, we analyze the Libri-Adapt dataset to demonstrate its feasibility in supporting unsupervised domain adaptation research for ASR models. To this end, we perform a number of experiments where we train an ASR model on a particular domain and evaluate its performance on another domain. More specifically, we use the Mozilla DeepSpeech2 pre-trained model~\cite{deepspeech} (release 0.5.0) as the base model which has a low word error rate (WER)  of 8.22\% on the LibriSpeech dataset. We fine-tune the base model on the training data from each domain and thereafter, the fine-tuned model for each domain is tested on the held-out test set from other domains to compute the WER.     

% \parjump
% \noindent
% \textbf{Quantifying the domain shift.} Firstly, we aim to quantify the domain shift across various domains in the dataset. In unsupervised domain adaptation, only unlabeled data is available from the target domain, therefore we employ Proxy A-Distance (PAD) ~\cite{} as a metric to quantify the shift between the source and target distributions. PAD, proposed by Ben-David et al. ~\cite{}, is a measure of the  $\mathcal{H}$-discrepancy between two distributions and is computed as 
% \begin{equation}
%   \text{PAD} (S, T) = 2 * ( 1 - 2 * e)
% \end{equation}

% where $e$ represents the error of a classifier trained to discriminate data from the source (S) and target (T) distributions. The lower the PAD, the closer two distributions are.   

% \todo{Describe the results}

\parjump
\noindent
\textbf{WER across Microphones.}  First, we analyze the impact of domain shift caused by microphone variability on the accuracy of ASR models. Table~\ref{tab:mic_asr} shows the WER of the DeepSpeech2 model when it is fine-tuned for different `training' microphones (in columns) and tested on other microphones (in rows). As expected, in most cases, it can be observed that when data from the same microphone is used for training and testing the model, it achieves the smallest WER, for e.g., 11.39\% in the case of PlayStation Eye. However, in cases of domain shift caused by microphone mismatch, the WER increases to 24.18\% when the PlayStation Eye model is tested on Matrix Voice. We also note that the WER obtained when models from other microphones are tested on Matrix Voice is significantly high (above 20\% in all cases), however when the model is fine-tuned on Matrix Voice, we can achieve a reasonable WER of 13.05\%. This highlights the opportunity for domain adaptation algorithms to adapt or align the speech feature representations across microphones when ASR models are deployed in scenarios with microphone mismatch.       

% \usepackage{booktabs}
\begin{table}[]
\centering
\scriptsize
\begin{tabular}{@{}|c|c|c|c|c|c|c|@{}}
\toprule
 & \multicolumn{6}{c|}{\textbf{Training Microphones}} \\ \midrule
 & \multicolumn{1}{c|}{\begin{tabular}[c]{@{}c@{}}Matrix \\ Voice\end{tabular}} & \multicolumn{1}{c|}{ReSpeaker} & \multicolumn{1}{c|}{USB} & \multicolumn{1}{c|}{Shule} & \multicolumn{1}{c|}{Nexus 6} & \multicolumn{1}{c|}{\begin{tabular}[c]{@{}c@{}}PlayStation\\ Eye\end{tabular}} \\ \midrule
\begin{tabular}[c]{@{}c@{}}Matrix\\ Voice\end{tabular} & 13.05 & 28.0 & 20.64 & 24.95 & 26.21 & 24.18  \\\midrule[0.5pt]
ReSpeaker & 16.06 & \textbf{12.66} & 14.49 & 16.09 & 15.93 & 15.21 \\\midrule[0.5pt]
USB & 13.30 & 14.44 & \textbf{11.40} & 14.41 & 15.39 & 15.38 \\ \midrule[0.5pt]
Shule & \textbf{12.58} & 13.10 & 11.61 & \textbf{10.19} & 12.96 & 13.07 \\\midrule[0.5pt]
Nexus 6 & 13.09 & 12.81 & 11.99 & 12.73 & \textbf{12.27} & 15.28 \\\midrule[0.5pt]
\begin{tabular}[c]{@{}c@{}}PlayStation\\ Eye\end{tabular} & 14.01 & 13.65 & 14.24 & 16.16 & 16.52 & \textbf{11.39} \\ \bottomrule
\end{tabular}
\caption{WER of a fine-tuned DeepSpeech2 model trained and tested on various microphone pairs for Clean, US English speech. The columns correspond to the training microphone domain and rows correspond to the test microphone domain.}
\vspace{-0.3cm}
\label{tab:mic_asr}
\end{table}

\begin{figure}[t]
    \centering
    \includegraphics[width=0.8\linewidth]{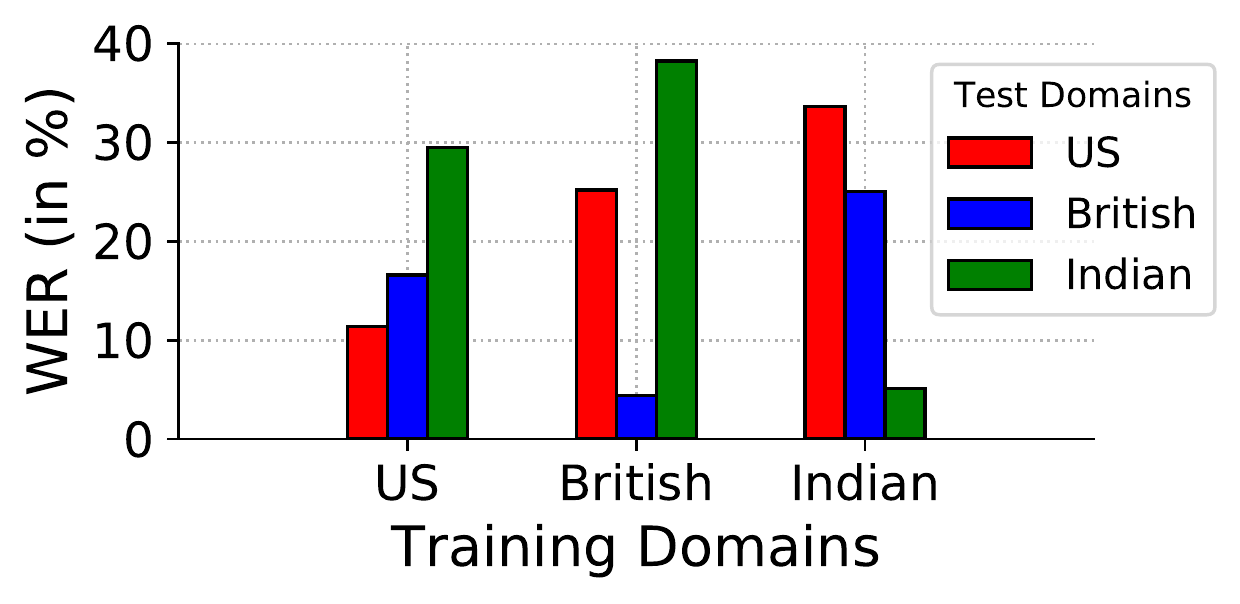}
    \vspace{-0.3cm}
    \caption{WER of a fine-tuned DeepSpeech2 model trained and tested on various accent datasets recorded on a \emph{ReSpeaker} microphone in a \emph{Clean} background.}
    \vspace{-0.3cm}
    \label{fig:accent_asr}
\end{figure}

\parjump
\noindent
\textbf{WER across Accents.} Next, we evaluate the impact of accent changes on ASR models. For this experiment, we set the microphone domain to \emph{ReSpeaker} and background noise to \emph{Clean}, which enables us to isolate the impact of accent variability on the model. As shown in Figure~\ref{fig:accent_asr}, there is a significant impact on WER caused by variations in speaker accents -- in particular, the generalizability of US and British English accent models to Indian English accent is quite poor, for instance, the WER of a model trained on US accent increases from 11.4\% to 29.5\% when it is tested on data with an Indian accent. Interestingly, we also observe that the WER of the British and Indian accent models on their own domains is very low ($\sim$5\%) as compared to the US accent model (11.4\%) -- a possible explanation of this result is that British and Indian accent datasets were generated by the Google WaveNet TTS model which supports a limited number (5-6) of user voices. As such, it was easier to fine-tune the model for these accents, as compared to US accent which had a total of 251 unique users. From a domain adaptation perspective, this represents a practical, albeit challenging, scenario: how to adapt a model trained on a small user group to a larger user group with different accents.

% % Please add the following required packages to your document preamble:
% % \usepackage{booktabs}
% \begin{table}[]
% \centering
% \begin{tabular}{@{}|c|c|c|c|c|@{}}
% \toprule
% \multicolumn{1}{|l|}{} & \multicolumn{4}{c|}{\textbf{Training Domains}} \\ \midrule
%  & Clean & Rain & Wind & Laughter \\ \midrule
% Clean & \textbf{11.70} & \textbf{19.07} & \textbf{16.22} & \textbf{13.84} \\ \midrule
% Rain & 66.89 & 32.01 & 50.40 & 54.70 \\ \midrule
% Wind & 40.17 & 38.26 & 21.21 & 35.03 \\ \midrule
% Laughter & 40.62 & 39.31 & 38.51 & 19.22 \\ \bottomrule
% \end{tabular}
% \caption{}
% \label{tab:noises}
% \end{table}

% \usepackage{booktabs}

\begin{table}[h]
\centering
\begin{tabular}{@{}|c|c|c|c|c|@{}}
\toprule
\multicolumn{1}{|l|}{} & \multicolumn{4}{c|}{\textbf{Training Domains}} \\ \midrule
 & Clean & Rain & Wind & Laughter \\ \midrule
Clean & \textbf{11.70} & \textbf{19.07} & {16.22} & \textbf{13.84} \\ \midrule
Rain & 30.32 & 21.63 & 23.45 & 23.56 \\ \midrule
Wind & 17.20 & 22.35 & \textbf{14.62} & 15.34 \\ \midrule
Laughter & 22.14 & 21.93 & 21.13 & 14.52 \\ \bottomrule
\end{tabular}
\caption{WER of a fine-tuned DeepSpeech2 model trained and tested on various background noise domains.}
\label{tab:noises}
\end{table}

% \begin{table}[]
% \centering
% \small
% \begin{tabular}{@{}llll@{}}
% \toprule
%  & US & British & Indian \\ \midrule
% US & \textbf{11.39} & 25.19 & 33.64 \\ \midrule
% British & 16.57 & \textbf{4.38} & 25.05 \\ \midrule
% Indian & 29.49 & 38.24 & \textbf{5.09} \\ \bottomrule
% \end{tabular}
% \caption{}
% \label{tab:my-table}
% \end{table}

\parjump
\noindent
\textbf{WER across Background Noise.} In Table~\ref{tab:noises}, we show the cross-domain results when ASR models are trained and tested on different background noise pairs. In this experiment, we set the microphone domain to USB and accent to US English, which helps in isolating the effect of background noise on model performance. The results in Table~\ref{tab:noises} confirm that domain shifts caused by ambient noise can degrade the performance of state-of-the-art ASR models, hence making it a pertinent problem for domain adaptation research.    

\parjump
\noindent
\textbf{Studying multi-domain shifts.} So far, we have analyzed the effect of a single form of domain shift on ASR models. However, Libri-Adapt can also be utilized to study the effect of multiple forms of domain shifts on ASR models. As shown in Table~\ref{tab:multi}, we can combine microphones, accents and noise conditions in various ways to form a domain; and then evaluate the cross-domain generalizability of ASR models. For instance, we observe in Table~\ref{tab:multi} that when a model trained on US accented clean speech collected on a ReSpeaker microphone (\emph{Domain 1}) is tested on Indian accented speech with Rain background noise recorded on a PlayStation Eye (\emph{Domain 2}), it suffers a WER increase of nearly 29.8\%. In summary, Libri-Adapt allows researchers to create various complex adaptation scenarios to test the generalizability of their domain adaptation algorithms.  

% \usepackage{booktabs}
\begin{table}[h]
\centering
\small
\begin{tabular}{@{}|c|c|c|c|c|@{}}
\toprule
 & \multicolumn{4}{c|}{\textbf{Training Domains}} \\ \midrule
 & Domain 1 & Domain 2 & Domain 3 & Domain 4 \\ \midrule
Domain 1 & \textbf{12.86} & 47.01 & 37.23 & 15.64 \\ \midrule
Domain 2 & 42.68 & \textbf{6.37} & 59.25 & 43.05 \\ \midrule
Domain 3 & 27.09 & 42.73 & \textbf{4.80} & 25.14 \\ \midrule
Domain 4 & 19.07 & 52.64 & 46.18 & \textbf{14.26} \\ \bottomrule
\end{tabular}
\caption{Libri-Adapt facilitates a variety of domain adaptation experiments by combining microphones, accents and noise conditions. This table shows the WER for various domain adaptation tasks. \emph{Domain 1} = ReSpeaker-US-Clean; \emph{Domain 2} = PlayStationEye-Indian-Rain; \emph{Domain 3} = Shule-British-Laughter; \emph{Domain 4} = USB-US-Wind}
\vspace{-0.4cm}
\label{tab:multi}
\end{table}

% \vspace{-0.3cm}
\section{Conclusion}
% \vspace{-0.2cm}
\label{sec:conclusion}
We presented a speech dataset to facilitate the study and development of unsupervised domain adaptation algorithms for ASR models. The dataset, derived from the Librispeech corpus, contains 7,200 hours of training data from 72 different domains spanning across 6 microphones, 3 speaker accents and 4 background noise conditions. One limitation of this work is that in order to create a large-scale and multi-domain dataset without incurring significant costs, we adopted the methodology of replay-and-record which may introduce minor speaker-related effects in the dataset.  As a future work, we will develop unsupervised domain adaptation algorithms to tackle the domain shifts in the dataset. The Libri-Adapt dataset is released publicly under the CC BY 4.0 license and can be accessed from \url{https://bit.ly/2UK1McG}. Updated baseline results with a newer release of DeepSpeech2 model (0.8.2) are also available at this URL.

% \begin{figure}[htb]

% \begin{minipage}[b]{1.0\linewidth}
%   \centering
%   \centerline{\includegraphics[width=8.5cm]{image1}}
% %  \vspace{2.0cm}
%   \centerline{(a) Result 1}\medskip
% \end{minipage}
% %
% \begin{minipage}[b]{.48\linewidth}
%   \centering
%   \centerline{\includegraphics[width=4.0cm]{image3}}
% %  \vspace{1.5cm}
%   \centerline{(b) Results 3}\medskip
% \end{minipage}
% \hfill
% \begin{minipage}[b]{0.48\linewidth}
%   \centering
%   \centerline{\includegraphics[width=4.0cm]{image4}}
% %  \vspace{1.5cm}
%   \centerline{(c) Result 4}\medskip
% \end{minipage}
% %
% \caption{Example of placing a figure with experimental results.}
% \label{fig:res}
% %
% \end{figure}

%\vfill
%\pagebreak

\bibliographystyle{IEEEbib}
\bibliography{strings,refs}

\end{document}